\documentclass[aps,pre,twocolumn,superscriptaddress,nofootinbib,longbibliography]{revtex4-1}

\usepackage{palatino}
\usepackage{amsmath}
\usepackage{amssymb}

\usepackage{graphicx}
\usepackage{dcolumn}
\usepackage{boxedminipage}
\usepackage{verbatim}
\usepackage[colorlinks=true,citecolor=blue,linkcolor=blue]{hyperref}

\graphicspath{{figures/}}

\newcommand{\bfv}[1]{{\mbox{\boldmath{$#1$}}}}
\newcommand{\bfm}[1]{{\bf #1}}          

\newcommand{\T}{\mathrm{T}}                                

\begin{document}

\title{Bayesian hidden Markov model analysis of single-molecule force spectroscopy: Characterizing kinetics under measurement uncertainty}

\author{John D. Chodera}
 \thanks{Corresponding author}
 \affiliation{California Institute of Quantitative Biosciences (QB3), University of California, Berkeley, CA 94720, USA}

\author{Phillip Elms}
 \affiliation{California Institute of Quantitative Biosciences (QB3), University of California, Berkeley, CA 94720, USA}
 \affiliation{Biophysics Graduate Group, University of California, Berkeley, CA 94720, USA}
 \affiliation{Jason L.~Choy Laboratory of Single Molecule Biophysics, Institute for Quantitative Biosciences, University of California, Berkeley, CA 94720, USA}

\author{Frank No\'{e}}
 \affiliation{DFG Research Center Matheon, FU Berlin, Arnimallee 6, 14195 Berlin, Germany}

\author{Bettina Keller}
 \affiliation{DFG Research Center Matheon, FU Berlin, Arnimallee 6, 14195 Berlin, Germany}

\author{Christian M. Kaiser}
 \affiliation{California Institute of Quantitative Biosciences (QB3), University of California, Berkeley, CA 94720, USA}
 \affiliation{Department of Physics, University of California, Berkeley, CA 94720, USA}

\author{Aaron Ewall-Wice}
 \affiliation{University of Chicago, IL 60637, USA}

\author{Susan Marqusee}
 \affiliation{California Institute of Quantitative Biosciences (QB3), University of California, Berkeley, CA 94720, USA}
 \affiliation{Department of Molecular \& Cell Biology, University of California, Berkeley, CA 94720, USA}
 \affiliation{Jason L.~Choy Laboratory of Single Molecule Biophysics, Institute for Quantitative Biosciences, University of California, Berkeley, CA 94720, USA}

\author{Carlos Bustamante}
 \affiliation{California Institute of Quantitative Biosciences (QB3), University of California, Berkeley, CA 94720, USA}
 \affiliation{Department of Molecular \& Cell Biology, University of California, Berkeley, CA 94720, USA}
 \affiliation{Jason L.~Choy Laboratory of Single Molecule Biophysics, Institute for Quantitative Biosciences, University of California, Berkeley, CA 94720, USA}
 \affiliation{Department of Physics, University of California, Berkeley, CA 94720, USA}
 \affiliation{Department of Chemistry, University of California, Berkeley, CA 94720, USA}
 \affiliation{Howard Hughes Medical Institute, University of California, Berkeley, CA 94720, USA}

\author{Nina Singhal Hinrichs}
 \affiliation{Departments of Statistics and Computer Science, University of Chicago, IL 60637, USA}

\date{\today}

\begin{abstract}

Single-molecule force spectroscopy has proven to be a powerful tool for studying the kinetic behavior of biomolecules.
Through application of an external force, conformational states with small or transient populations can be stabilized, allowing them to be characterized and the statistics of individual trajectories studied to provide insight into biomolecular folding and function.
Because the observed quantity (force or extension) is not necessarily an ideal reaction coordinate, individual observations cannot be uniquely associated with kinetically distinct conformations.
While maximum-likelihood schemes such as hidden Markov models have solved this problem for other classes of single-molecule experiments by using temporal information to aid in the inference of a sequence of distinct conformational states, these methods do not give a clear picture of how precisely the model parameters are determined by the data due to instrument noise and finite-sample statistics, both significant problems in force spectroscopy.
We solve this problem through a \emph{Bayesian} extension that allows the experimental uncertainties to be directly quantified, and build in detailed balance to further reduce uncertainty through physical constraints.
We illustrate the utility of this approach in characterizing the three-state kinetic behavior of an RNA hairpin in a stationary optical trap.


\end{abstract}

\maketitle

\small


\section{Introduction}
\label{section:introduction}

Recent advances in biophysical measurement have led to an unprecedented ability to monitor the dynamics of single biological macromolecules, such as proteins and nucleic acids~\cite{ritort:j-phys:2006:single-molecule-review}.
As a new approach to probing the behavior of biological macromolecules, these experiments promise to change the way we study folding, dynamics, catalysis, association, transcription, translation, and motility, providing otherwise-inaccessible information about microscopic kinetics, energetics, mechanism, and the stochastic heterogeneity inherent in these processes.
Advances in instrumentation for optical force spectroscopy in particular have produced instruments of extraordinary stability, precision, and temporal resolution~\cite{block:rev-sci-instrum:2004:optical-trapping,bustamante:annu-rev-biochem:2008:optical-tweezers} that have already demonstrated great utility in the study of biomolecules in the presence of externally perturbative forces~\cite{bustamante:nature:2003:dna-mechanics,block:curr-opin-chem-biol:2008:rna,bustamante:cell:2011:central-dogma}. 
Under external force, it becomes possible to stabilize and characterize short-lived conformational states, such as protein folding and unfolding intermediates~\cite{clarke:jmb:2002:titin-afm,cecconi-shank-bustamante-marqusee:science:2005:rnase-h,rief:pnas:2010:pmf}.

\begin{figure}[tbp]
\noindent
\resizebox{\columnwidth}{!}{\includegraphics{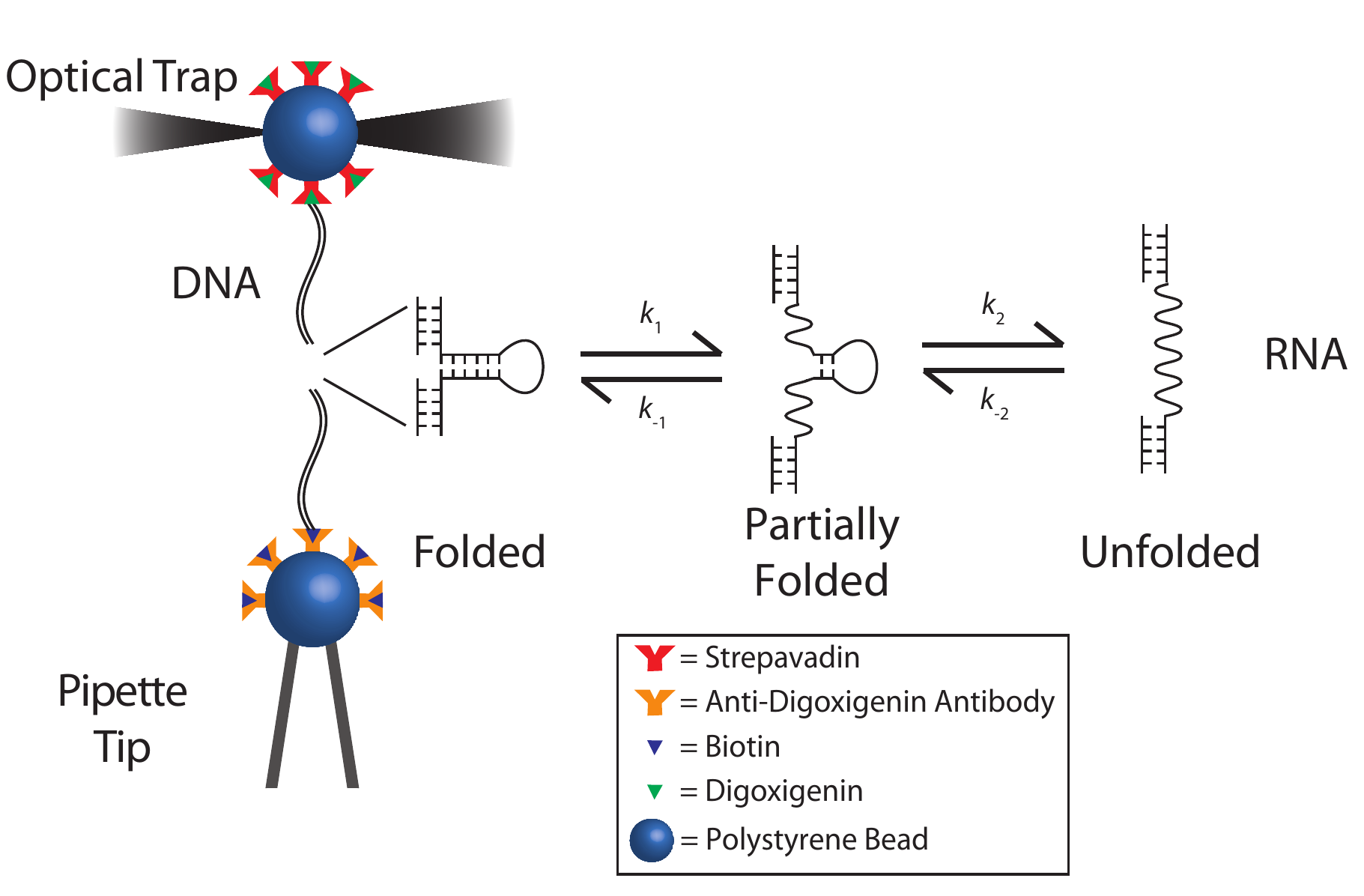}}
\caption{{\bf Single-molecule optical trapping configuration.}
The biomolecule of interest---here, the p5ab RNA hairpin---is tethered to two polystyrene beads by dsDNA handles.
The fluctuating force on one bead held in an optical trap is monitored, while the other bead is held suctioned to a micropipette tip.
Conformational transitions of the hairpin---such as transitions among the three kinetically metastable states illustrated here---are observed indirectly through motion of the bead in the trap.
\label{figure:trap-configuration}}
\end{figure}

In a typical single-molecule optical trapping experiment, a protein or nucleic acid is tethered to two polystyrene beads by dsDNA handles that prevent the molecule under study from interacting with the beads (see Figure~\ref{figure:trap-configuration}).
The handle-biomolecule-handle assembly---referred to as a \emph{fiber}---is associated with the beads through tight noncovalent interactions, with one bead held in an optical trap and the other either suctioned to a micropipette or held in a second optical trap.
During an experiment, the position of the bead within the laser trap is monitored, and either the relative displacement from the trap center or the total force on the bead is recorded, resulting in a timeseries such as the one depicted in Figure~\ref{figure:model-stateassignments}.
The instrument can generally be operated in several modes: a \emph{force ramp} mode, in which the trap is translated rapidly enough to potentially carry the system out of equilibrium; an equilibrium \emph{passive} mode, in which the trap is held fixed; and a \emph{constant force-feedback} mode, in which the trap is continually repositioned to maintain a set constant force on the fiber. 
Here, we concern ourselves with the latter two classes of experiment, though nonequilibrium experiments remain an exciting topic of active research~\cite{ritort:adv-chem-phys:2008:nonequilibrium-review}.

Often, the dynamics observed in these experiments appears to be dominated by stochastic transitions between two or more strongly metastable conformational states~\cite{schuette-huisinga:2002:biomolecular-conformations-as-metastable-states,chodera-singhal:jcp:2007:automatic-state-decomposition}---regions of conformation space in which the system remains for long times before making a transition to another conformational state.
These transitions are generally well-described by first-order kinetics~\cite{noe:pnas:2009:ww-domain}.
While visual inspection of the dynamics may suggest the clear presence of multiple metastable states, quantitative characterization of these states is often difficult.
First, the observed force or extension is unlikely to correspond to a true reaction coordinate easily able to separate all metastable states~\cite{best-paci-hummer-dudko:jpcb:2008:pulling-coordinate,rief-paci:pre:2010:multidimensional-landscape,thirumalai:2011:prl:pfold,chodera-pande:arxiv:2011:single-molecule-pfold}, and second, measurement noise may further broaden the force or extension signatures of individual states, increasing their overlap.
Attempting to separate these states by simply dividing the observed force or extension range into regions, following current practice~\cite{woodside:science:2006:dna-hairpin-optical-trap,ritort:biophys-j:2011:short-handles}, can often lead to a high degree of state mis-assignment that results in the estimated rate constants and state distributions containing a significant amount of error~\cite{chodera:rate-theory} (see \emph{Supplementary Material: Comparison with threshold model}).

Hidden Markov models (HMMs)~\cite{rabiner:ieee-proceedings:1989:hmm-tutorial}, which use temporal information in addition to the instantaneous value of the observable (force or extension) to determine which conformational states the system has visited during the experiment, have provided an effective solution to the hidden state problem in many other classes of single-molecule experiments, such as ion channel currents~\cite{becker:pfluegers-arch:1994:ion-channel-hmm,sachs:biophys-j:2000:ion-channel-hmm,degunst-kuensch-schouten:j-am-stat-assoc:2001:ion-channel-hmm,qin:biophys-j:2004:ion-channel-hmm}, single-molecule FRET~\cite{andrec-levy-talaga:jpca:2003:photon-arrival-hmm,mckinney-joo-ha:biophys-j:2006:hmm-fret,lee:jpcb:2009:hmm-fret,scherer-dinner:jpcb:2009:rna-fret-hmm,gopich-szabo:jpcb:2009:decoding-fret}, and the stepping of motor proteins~\cite{smith:biophys-j:2001:hmm-actomyosin,sachs:biophys-j:2006:motor-hmm,sigworth:biophys-j:2010:hmm-motors}.
In applying hidden Markov modeling to the analysis of single-molecule force spectroscopy data, the observed force or extension trace is assumed to come from a realization of an underlying Markov chain, where the system makes history-independent transitions among a set of discrete conformational states with probabilities governed by a transition or rate matrix.
Data, in the form of force or bead-to-bead extension measurements, is sampled at an interval that ensures that sequential observations satisfy the Markov property of history-independence, though the appropriate interval depends on the properties of the experimental configuration.
Under a given set of external force conditions, each state has a distribution of forces or extensions associated with it.
Given observed timeseries data for forces or extensions, the maximum likelihood estimate (MLE) of the model parameters (transition rates and state force or extension distributions) and sequence of hidden states corresponding to the observed data can be determined by standard methods~\cite{baum:1970:ann-math-statist:baum-welch,viterbi:1967:IEEE-trans-info-theory:viterbi-algorithm}, as demonstrated in recent work~\cite{kruithof-vannoort:biophys-j:2009:hmm-force}.

Unfortunately, this approach has a number of significant drawbacks.
Due to technical limitations, experiments often suffer from limited statistics---the events of interest (transitions between states or visits to rare states) may occur only a few times during the course of the measurement, and data for additional fibers is time-consuming to collect.
As a result, while the MLE yields the most \emph{likely} set of model parameters, there may be enormous uncertainty in some of these parameters, and the uncertainty in multiple parameters may be correlated in complex nonlinear ways.
While methods exist for estimating the standard error or confidence intervals from \mbox{MLHMMs}~\cite{aittokallio-uusipaikka:technical-report:2000:hmm-standard-error}, these schemes can be prohibitively costly for long traces, and may still significantly underestimate the statistical error for short traces due to the normally-distributed error approximation inherent in the approach.
The high cost (both in terms of instrument and experimenter time) of collecting additional data also means that it is not a simple task to judge \emph{how much} data need be collected to test a particular hypothesis in a statistically meaningful way.
Worse yet, the standard algorithms employed to find the MLE may not even find the true maximum likelihood solution, instead converging to a local maximum in likelihood that is far from optimal~\cite{merialdo:ieee-trans-speech-audio-proc:1993:baum-welch-localit}.

Here, we resolve this issue through the use of a \emph{Bayesian} extension of hidden Markov models~\cite{robert:stat-prob-lett:1993:bayesian-hmm,chib:j-econometrics:1996:bayesian-hmm,scott:j-am-stat-assoc:2002:bayesian-hmm,ryden:bayesian-analysis:2008:bayesian-hmm} applicable to single molecule force experiments.
By sampling over the posterior distribution of model parameters and hidden state assignments instead of simply finding the most likely values, the experimenter is able to accurately characterize the correlated uncertainties in both the model parameters (transition rates and state force or extension distributions) and hidden state sequences corresponding to observed data.
Additionally, prior information (either from additional independent measurements or physical constraints) can be easily incorporated.
We also include a reversibility constraint on the transition matrix---in which microscopic detailed balance is imposed on the kinetics, as dictated by the physics of equilibrium systems~\cite{van-kampen}---which has been shown to significantly reduce statistical uncertainties in data-poor conditions~\cite{noe:jcp:2008:transition-matrix-sampling,metzner-noe-schuette:pre:2009:transition-matrix-sampling}.
The framework we present is based on Gibbs sampling~\cite{geman-geman:ieee-trans-pattern-analysis:1984:gibbs-sampling,jun-s-liu:mcmc}, allowing simple swap-in replacement of models for observable distributions, extension to multiple observables, and alternative models for state transitions.
Additionally, the Bayesian method provides a straightforward way to model the statistical outcome and assess the utility of additional experiments given some preliminary data, allowing the experimenter a powerful tool for assessing whether the cost of collecting additional data is outweighed by their benefits.
A Matlab implementation of this approach is available online [\url{http://simtk.org/home/bhmm}].


\section{Hidden Markov models for force spectroscopy}
\label{section:theory}

Suppose the temporal behavior of some observable $O(x)$ that is a function of molecular configuration $x$---here, generally force or molecular extension---is observed at temporal intervals $\Delta t$ to produce a timeseries $o_t$, where $t = 0, 1, \ldots, L$.
An instantaneous observation $o_t$ does not necessarily contain enough information to unambiguously identify the current conformational state the molecule occupies; to infer the hidden state, we must also make use of the temporal information in the observed trace.
We restrict ourselves to consideration of scalar functions $O(x)$, but the generalization to multidimensional probes (or multiple probes, such as combined force and fluorescence measurements~\cite{chemla:nature-methods:2011:fluorescence-force}) and multiple observed temporal traces is straightforward.

We presume the system under study has $M$ kinetically distinct states, in the sense that the system generally remains in a given state for several observation intervals $\Delta t$, but these states may not necessarily represent highly populated states of the system at equilibrium.
We treat these conformational states as the \emph{hidden states} of the model, because we cannot \emph{directly} observe the identity of the metastable state in which the system resides.
The hidden Markov model presumes the observed data $\bfm{O} \equiv \{o^{}_t\}$ was generated according to the following model dependent on parameters $\bfm{\Theta} \equiv \{\bfm{T}, \bfm{E}\}$, where $\bfm{T}$ is an $M \times M$ row-stochastic transition matrix and $\bfm{E}$ a set of emission parameters governing the observable (force or extension) distributions for each of the $M$ hidden states, and prior information about the initial state distribution $\bfv{\rho}^{}$,
\begin{eqnarray}
\mathbb{P}(s^{}_0) &=& \rho^{}_{s^{}_0} \nonumber \\
\mathbb{P}(s^{}_t \mid s^{}_{t-1}, \bfm{T}) &=& T_{s^{}_{t-1} s^{}_{t}} , \:\: t \ge 1 \nonumber \\
\mathbb{P}(o^{}_t \mid s^{}_t, \bfm{e}_{s^{}_{t}}) &=& \varphi(o^{}_{t} \mid \bfm{e}_{s^{}_{t}}) \label{equation:hmm-process} .
\end{eqnarray}
In diagrammatic form, the observed state data $\{o_t\}$ and corresponding hidden state history $\{s_t\}$ can be represented 
\begin{eqnarray}
\begin{array}{llllllllll}
 \stackrel{\bfv{\rho}}{\longrightarrow} & s^{}_{0} & \stackrel{\bfm{T}}{\longrightarrow} & s^{}_{1} & \stackrel{\bfm{T}}{\longrightarrow} & s^{}_{2} & \stackrel{\bfm{T}}{\longrightarrow} & \cdots & \stackrel{\bfm{T}}{\longrightarrow} & s^{}_{L^{}} \\
 & \downarrow \varphi &  & \downarrow \varphi & & \downarrow \varphi & & & & \downarrow \varphi \\
 & o^{}_{0} & & o^{}_{1} & & o^{}_{2} &  & & & o^{}_{L^{}} \\
\end{array}
\end{eqnarray} 
The initial state distribution $\bfv{\rho}^{}$  reflects our knowledge of the initial conditions of the experiment that collected data $\bfm{o}^{}$.
In the case that the experiment was prepared in equilibrium, $\bfv{\rho}$ corresponds to the equilibrium distribution $\bfv{\pi}$ of the model transition matrix $\bfm{T}$; if the experiment was prepared out of equilibrium, $\bfv{\rho}$ may be chosen to reflect some other prior distribution (e.g.~the uniform prior).

State transitions $(s^{}_{t-1} \rightarrow s^{}_{t})$ are governed by the discrete transition probability $T_{s^{}_{t-1} s^{}_{t}}$.
The \emph{Markov property} of HMMs prescribes that the probability that a system originally in state $i$ at time $t$ is later found in state $j$ at time $t+1$ is dependent only on knowledge of the state $i$, and given by the corresponding matrix element $T_{ij}$ of the (row-stochastic) transition matrix $\bfm{T}$.
Alternatively, one could instead use the rate matrix $\bfm{K}$, related to the transition matrix $\bfm{T}$ through the equation $\bfm{T} = e^{\bfm{K} \Delta t}$.  
If the processes described by $\bfm{T}$ or $\bfm{K}$ are slow compared to the observation interval $\Delta t$, then we can easily estimate the rate matrix from the associated transition matrix in a way that avoids the matrix logarithm, through the expansion $\bfm{K} \approx (\bfm{T} - \bfm{I}) / \Delta t$, where $\bfm{I}$ denotes the $M \times M$ identity matrix.

The probabilistic ``emission'' of observables from each state $(s^{}_{t} \rightarrow o^{}_{t})$ is governed by the continuous emission probability $\varphi(o^{}_{t} \mid \bfm{e}_{s^{}_{t}})$, parametrized by observable \emph{emission} parameters $\bfm{e}$.
For example, in the force spectroscopy applications described here, $\varphi(o \mid \bfm{e}_s)$ is taken to be a univariate normal (Gaussian) distribution parameterized by a mean $\mu$ and variance $\sigma^2$ that characterize each state, such that $\bfm{e}_i \equiv \{\mu_i, \sigma^2_i\}$.
Other choices of observable distribution can easily be substituted in a modular way without affecting the structure of the algorithms presented here.

Given the HMM process specified in Eq.~\ref{equation:hmm-process}, the probability of observing data $\bfm{O}$ given the model parameters $\bfm{\Theta}$ is then,
\begin{eqnarray}
P(\bfm{O} \mid \bfm{\Theta}) &=& \sum_{\bfm{S}}  \rho_{s^{}_{0}} \varphi(o^{}_{0} \mid \bfm{e}_{s^{}_{0}}) \prod_{t=1}^{L^{}} T_{s^{}_{t-1} s^{}_{t}} \varphi(o^{}_{t} \mid \bfm{e}_{s^{}_{t}}) \label{equation:hmm-likelihood} ,
\end{eqnarray}
where the sum over hidden state histories $\bfm{S}$ is shorthand for
\begin{eqnarray}
\sum_{\bfm{S}} \equiv \sum_{s^{}_{0}=1}^M \sum_{s^{}_{1}=1}^M \cdots \sum_{s^{}_{L^{}}=1}^M .
\end{eqnarray}
If multiple independent traces $\{o_t\}$ are available, the probability $P(\bfm{O} \mid \bfm{\Theta})$ is simply the product of Eq.~\ref{equation:hmm-likelihood} for the independent traces.

\subsection{Maximum likelihood hidden Markov model (MLHMM)}

The standard approach to construct an HMM from observed data is to compute the \emph{maximum likelihood estimator} (MLE) for the model parameters $\bfm{\Theta} \equiv \{\bfm{T}, \bfm{E}\}$, which maximize the probability of the observed data $\bfm{O}$ given the model,
\begin{eqnarray}
\hat{\bfm{\Theta}} &=& \arg\max_\bfm{\Theta} P(\bfm{O} \mid \bfm{\Theta}) ,
\end{eqnarray}
yielding MLE estimates of transition matrix $\hat{\bfm{T}}$ and state emission parameters $\hat{\bfm{E}}$.
Typically, determination of the model parameters $\bfm{\Theta}$ is carried out using the Baum-Welch algorithm~\cite{baum:1970:ann-math-statist:baum-welch}.

Once the MLE parameters $\hat{\bfm{\Theta}}$ are determined, the most likely hidden state history that produced the observations $\bfm{O}$ can be determined using these parameters:
\begin{eqnarray}
\hat{\bfm{S}} &=& \arg\max_\bfm{S} P(\bfm{S} \mid \bfm{O}, \hat{\bfm{\Theta}}) .
\end{eqnarray}
This is typically carried out using the Viterbi algorithm~\cite{viterbi:1967:IEEE-trans-info-theory:viterbi-algorithm}, a classic example of dynamic programming.


\subsection{Bayesian hidden Markov model (BHMM)}

Instead of simply determining the model that maximizes the likelihood of observing the data $\bfm{O}$ given the model parameters $\bfm{\Theta}$, we can make use of Bayes' theorem to compute the \emph{posterior} distribution of model parameters given the observed data:
\begin{eqnarray}
P(\bfm{\Theta} \mid \bfm{O}) &\propto& P(\bfm{O} \mid \bfm{\Theta}) P(\bfm{\Theta}) \label{equation:bayes-theorem}.
\end{eqnarray}
Here, $P(\bfm{\Theta})$ denotes a \emph{prior} distribution that encodes any \emph{a priori} information we may have about the model parameters $\bfm{\Theta}$.
This prior information might include, for example, physical constraints (such as ensuring the transition matrix satisfies detailed balance) or prior rounds of inference from other independent experiments.

Making use of the likelihood (Eq.~\ref{equation:hmm-likelihood}), the model posterior is then given by,
\begin{eqnarray}
P(\bfm{\Theta} \mid \bfm{O}) &\propto& P(\bfm{\Theta}) \sum_{\bfm{S}}\rho_{s^{}_{0}} \varphi(o^{}_{0} \mid \bfm{e}_{s^{}_{0}}) \prod_{t=1}^{L^{}} T_{s^{}_{t-1} s^{}_{t}} \varphi(o^{}_{t} \mid \bfm{e}_{s^{}_{t}})  \label{equation:full-posterior} .
\end{eqnarray}
Drawing samples of $\bfm{\Theta}$ from this distribution will, in principle, allow the \emph{confidence} with which individual parameters and combinations thereof are known, given the data (and subject to the validity of the model of Eq.~\ref{equation:hmm-process} in correctly representing the process by which the observed data is generated).
While the posterior $P(\bfm{\Theta}|\bfm{O})$ is complex, we could in principle use a Markov chain Monte Carlo (MCMC) approach~\cite{jun-s-liu:mcmc} to sample it.
In its current form, however, this would be extremely expensive due to the sum over all hidden state histories $\bfm{S}$ appearing in ratios involving Eq.~\ref{equation:full-posterior}.
Instead, we introduce the hidden state histories $\bfm{S}$ as an auxiliary variable, sampling from the augmented posterior,
\begin{eqnarray}
P(\bfm{\Theta}, \bfm{S} \mid \bfm{O}) &\propto& \left[ \rho_{s^{}_{0}} \varphi(o^{}_{0} \mid \bfm{e}_{s^{}_{0}}) \prod_{t=1}^{L^{}} T_{s^{}_{t-1} s^{}_{t}} \varphi(o^{}_{t} \mid \bfm{e}_{s^{}_{t}}) \right] P(\bfm{\Theta}) . \nonumber \\\label{equation:augmented-posterior} 
\end{eqnarray}
which makes it much less costly to compute the ratios required for MCMC on the augmented $(\bfm{\Theta},\bfm{S})$ parameter space.

If we presume the prior is separable, such that $P(\bfm{\Theta}) \equiv P(\bfm{T}) P(\bfm{E})$, we can sample from the augmented posterior (Eq.~\ref{equation:augmented-posterior}) using the framework of \emph{Gibbs sampling}~\cite{jun-s-liu:mcmc}, in which the augmented model parameters are updated by sampling from the conditional distributions,
\begin{eqnarray}
P(\bfm{S} \mid \bfm{T}, \bfm{E}, \bfm{O}) &\propto& \rho_{s^{}_{0}} \varphi(o^{}_{0} \mid \bfm{e}_{s^{}_{0}}) \prod_{t=1}^{L^{}} T_{s^{}_{t-1} s^{}_{t}} \varphi(o^{}_{t} \mid \bfm{e}_{s^{}_{t}}) \nonumber \\
P(\bfm{T} \mid \bfm{E}, \bfm{S}, \bfm{O}) &=& P(\bfm{T} \mid \bfm{S}) \propto P(\bfm{T}) \prod_{t=1}^{L^{}} T_{s^{}_{t-1} s^{}_{t}} \nonumber \\
P(\bfm{E} \mid \bfm{S}, \bfm{T}, \bfm{O}) &=& P(\bfm{E} \mid \bfm{S}, \bfm{O}) \propto P(\bfm{E}) \prod_{t=0}^{L^{}} \varphi(o^{}_{t} \mid \bfm{e}_{s^{}_{t}}) \label{equation:conditional-gibbs} .
\end{eqnarray}
The equalities on the second and third lines reflect the conditional independence of the hidden Markov model defined by Eq.~\ref{equation:hmm-process}.
When only the model parameters $\bfm{\Theta} \equiv \{ \bfm{T}, \bfm{E} \}$ or the hidden state histories $\bfm{S}$ are of interest, we can simply marginalize out the uninteresting variables by sampling from the augmented joint posterior for $\{\bfm{T}, \bfm{E}, \bfm{S}\}$ and examine only the variables of interest.
In addition, the structure of the Gibbs sampling scheme above allows individual components (such as the observable distribution model $\varphi(o \mid \bfm{e})$ or transition probability matrix $\bfm{T}$) to be modified without affecting the structure of the remainder of the calculation.

In the illustrations presented here, we employ a Gaussian observable distribution model for $\varphi(o \mid \bfm{e})$,
\begin{eqnarray}
\varphi(o \mid \bfm{e}) &=& \varphi(o \mid \mu, \sigma^2) = \frac{1}{\sqrt{2 \pi} \sigma} \exp\left[-\frac{1}{2}\frac{(o - \mu)^2}{\sigma^2} \right] \label{equation:gaussian-observable} ,
\end{eqnarray}
where $\mu$ is the mean force or extension characterizing a particular state, and $\sigma$ is the standard deviation or width of forces or extensions corresponding to that state.
We note that marginal posterior distributions of each mean $P(\mu_i | \bfm{O})$ reflect the statistical uncertainty in how well the mean force or position is determined, and need not correspond to the standard deviation $\sigma_i$, which may be much broader (or narrower, depending on the situation).


\section{Algorithms}
\label{section:algorithms}

\subsection{Generating an initial model}

To initialize either computation of the MLHMM or sampling from the posterior for the BHMM, an initial model that respects any constraints imposed in the model prior $P(\bfm{\Theta})$ must be selected.
Here, we employ a Gaussian observable distribution model for $\varphi(o \mid \bfm{e})$ (Eq.~\ref{equation:gaussian-observable}) and enforce that the transition matrix $\bfm{T}$ satisfy detailed balance.

\subsubsection{Observable parameter estimation}

We first initialize the observed distributions of each state by fitting a Gaussian mixture model with $M$ states to the pooled observed data $\bfm{O}$, ignoring temporal information:
\begin{eqnarray}
P(\bfm{O} \mid \bfv{\pi}, \bfm{E}) &=& \prod_{t=0}^{L^{}} \sum_{m=1}^M \pi_m \varphi(o^{}_{t} \mid \mu_m, \sigma^2_m) ,
\end{eqnarray}
where the state observable emission probability vector $\bfm{E} \equiv \{\bfm{e}_1,\ldots,\bfm{e}_M\}$ and $\bfm{e}_m \equiv \{\mu_m, \sigma_m^2\}$ with $\mu_m$ denoting the observable mean and $\sigma_m^2$ the variance for state $m$ for the Gaussian mixture model.
The vector $\bfv{\pi}$ is composed of equilibrium state populations $\{ \pi_1, \ldots , \pi_M\}$ with $\pi_m \ge 0$ and $\sum_{m=1}^M \pi_m = 1$.

A first approximation to $\bfv{\pi}$ and $\bfm{E}$ is computed by pooling and sorting the observed $o^{}_{t}$, and defining $M$ indicator functions $h_m(o)$ that separate the data into $M$ contiguous regions of the observed range of $o$ of roughly equal population.
Let $N_m \equiv \sum_{t=0}^{L^{}} h_m(o^{}_{t})$ denote the total number of observations falling in region $m$, and $N_\mathrm{tot} = \sum_{m=1}^M N_m$.
The initial parameters are then computed as,
\begin{eqnarray}
\pi_m &=& N_m / N_\mathrm{tot} \nonumber \\
\mu_m &=& N_m^{-1} \sum_{t=0}^{L^{}} o^{}_{t} \, h_m(o^{}_{t}) \\
\sigma^2_m &=& N_m^{-1} \sum_{t=0}^{L^{}} (o^{}_{t} - \mu_m)^2 \, h_m(o^{}_{t}) .
\end{eqnarray}

This approximation is then improved upon by iterating the expectation-maximization procedure described by Bilmes~\cite{bilmes:1998:expectation-maximization},
\begin{eqnarray}
\pi'_m &=& N_\mathrm{tot}^{-1} \sum_{t=0}^{L^{}} \chi_m(o^{}_t, \bfm{E}, \bfv{\pi}) \nonumber \\
\mu'_m &=& (\pi'_m N_\mathrm{tot})^{-1} \sum\limits_{t=0}^{L^{}} o^{}_{t} \, \chi_m(o^{}_{t}, \bfm{E}, \bfv{\pi}) \nonumber \\
{\sigma'}^2_m &=& (\pi'_m N_\mathrm{tot})^{-1} \sum\limits_{t=0}^{L^{}} (o^{}_{t} - \mu'_m)^2 \, \chi_m(o^{}_{t}, \bfm{E}, \bfv{\pi}) 
\end{eqnarray}
where the function $\chi_m(o, \bfm{E}, \bfv{\pi})$ is given by the fuzzy membership function,
\begin{eqnarray}
\chi_m(o, \bfm{E}, \bfv{\pi}) &=& \frac{\pi_m \, \varphi(o \mid \bfm{e}_m)}{\sum\limits_{l=1}^M \pi_l \, \varphi(o \mid \bfm{e}_l)} .
\end{eqnarray}
The iterative procedure is terminated at iteration $j$ when the change in the parameters $\{\bfv{\pi}, \bfv{\mu}, \bfv{\sigma}^2\}$ falls below a certain relative threshold, such as $\|\bfv{\pi}^{[j]} - \bfv{\pi}^{[j-1]}\|_2 / \|\bfv{\pi}^{[j]}\|_2 < 10^{-4}$.

\subsubsection{Transition matrix estimation}

Once initial state observable emission parameters $\bfm{E}$ are determined, an initial transition matrix is estimated using an iterative likelihood maximization approach that enforces detailed balance~\cite{noe:jcp:2011:msm-review}.
First, a matrix of fractional transition counts $\bfm{C} \equiv (c_{ij})$ is estimated using the membership function:
\begin{eqnarray}
c_{ij} &=& \sum_{t=1}^{L^{}} \chi_i(o^{}_{t-1}, \bfm{E}, \bfv{\pi}) \, \chi_j(o^{}_{t}, \bfm{E}, \bfv{\pi})
\end{eqnarray}
A symmetric $M \times M$ matrix $\bfm{X} \equiv (x_{ij})$ is initialized by
\begin{eqnarray}
x_{ij} = x_{ji} &=& c_{ij} + c_{ji} .
\end{eqnarray}
The iterative procedure described in Algorithm 1 of \cite{noe:jcp:2011:msm-review} is then applied.
For each update iteration, we first update the diagonal elements of $\bfm{X}$:
\begin{eqnarray}
x'_{ii} = \frac{c_{ii} (x_{i*} - x_{ii})}{c_{i*} - c_{ii}}  \:\:;\;\; c_{i*} = \sum_{j=1}^M c_{ij} \:\:;\:\: x_{i*} = \sum_{j=1}^M x_{ij} ,
\end{eqnarray}
followed by the off-diagonal elements:
\begin{eqnarray}
x'_{ij} = x'_{ji} &=& \frac{-b + \sqrt{b^2 - 4 a c}}{2 a}
\end{eqnarray}
where the quantities $a$, $b$, and $c$ are computed from $\bfm{X}$ and $\bfm{C}$,
\begin{eqnarray}
a &\equiv& c_{i*} - c_{ij} + c_{j*} - c_{ji} \nonumber \\
b &\equiv& c_{i*} (x_{j*} - x_{ji}) + c_{j*} (x_{i*} - x_{ij}) \nonumber \\
&&\mbox{}- (c_{ij} + c_{ji}) (x_{i*} - x_{ij} + x_{j*} - x_{ji}) \nonumber \\
c &\equiv& - (c_{ij} + c_{ji}) (x_{i*} - x_{ij}) (x_{j*} - x_{ji}) .
\end{eqnarray}
Once a sufficient number of iterations $j$ have been completed to compute a stable estimate of $\bfm{X}$ (such as the relative convergence criteria $\|\bfm{X}^{[j]} - \bfm{X}^{[j-1]}\|_2 / \|\bfm{X}^{[j]}\|_2 < 10^{-4}$, the maximum likelihood transition matrix estimate $\bfm{T}$ is computed as
\begin{eqnarray}
T_{ij} &=& x_{ij} / x_{i*} . 
\end{eqnarray}
Note that the equilibrium probability vector $\bfv{\pi}$ computed during the Gaussian mixture model fitting is not respected during this step.

\subsection{Fitting a maximum likelihood HMM}
\label{section:fitting-mlhmm}

The HMM model parameters $\bfm{\Theta} \equiv \{ \bfm{T}, \bfm{E} \}$ are fit to the observed data $\bfm{O}$ through use of the expectation-maximization (EM) algorithm~\cite{dempster-laird-rubin:1977:j-royal-statist-soc-b:em-algorithm}.
This is an iterative procedure, where the model parameters are subsequently refined through successive iterations.
The initial HMM is usually quick to compute, and can give the experimenter a rough idea of the model parameters, as well as providing a useful starting point for sampling models from the Bayesian posterior.

During each iteration, the Baum-Welch algorithm~\cite{baum:1970:ann-math-statist:baum-welch} is used to compute $\bfm{\Xi}^{} \equiv (\xi^{}_{tij})$, which represents the probability that the system transitions from hidden state $i$ at time $t-1$ to hidden state $j$ at time $t$, and $\gamma^{}_{ti}$, the probability that the system occupied state $i$ at time $t$.
This is accomplished by first executing the \emph{forward algorithm},
\begin{eqnarray}
\alpha_{tj} &=& \begin{cases}
\rho_j \, \varphi(o_0 \mid \bfm{e}_j) & t = 0 \\
\varphi(o_t \mid \bfm{e}_j) \sum_{i=1}^M \alpha_{(t-1)i} T_{ij} & t = 1,\ldots,L
\end{cases}
\end{eqnarray}
followed by the \emph{backward algorithm},
\begin{eqnarray}
\beta_{ti} &=& \begin{cases}
1 & t = L \\
\sum_{j=1}^M T_{ij} \varphi(o_{t+1} \mid \bfm{e}_j) \beta_{(t+1)j} & t = (L-1),\ldots,0
\end{cases}
\end{eqnarray}
The $L \times M \times M$ matrix $\bfm{\Xi}$ is then computed for $t = 0,\ldots,(L-1)$ as,
\begin{eqnarray}
\xi_{tij} &=& \alpha_{ti} \varphi(o_{t+1} \mid \bfm{e}_i) T_{ij} \beta_{(t+1)j} / \sum_{i=1}^M \alpha_{Ti} \\
\gamma_{ti} &=& \sum_{j=1}^M \xi_{tij}
\end{eqnarray}
In practice, the logarithms of these quantities are computed instead to avoid numerical underflow.

The aggregate matrix of expected transition counts $\bfm{C}~\equiv~(c_{ij})$ is then computed from $\bfm{\Xi}^{}$ as,
\begin{eqnarray}
c_{ij} &=& \sum_{t=0}^{L^{}-1} \xi_{tij}^{} .
\end{eqnarray}
This count matrix is used to update the maximum-likelihood transition matrix $\bfm{T}$ using the method of Prinz et al.~\cite{noe:jcp:2011:msm-review} described in the previous section.

The state observable distribution parameters $\bfm{E}$ are then updated from the $\gamma_{ti}$.
For the univariate normal distribution applied to force spectroscopy data here, we update the mean $\mu_i$ and variance $\sigma^2_i$ for state $i$ using the scheme,
\begin{eqnarray}
\mu'_i = \frac{\sum\limits_{t=0}^{L} o^{}_{t} \gamma^{}_{ti}}{\sum\limits_{t=0}^{L^{}} \gamma^{}_{ti}} \:\:;\:\: {\sigma'}^2_i = \frac{\sum\limits_{t=0}^{L^{}} (o^{}_{t} - \mu'_i)^2 \gamma^{}_{ti}}{\sum\limits_{t=0}^{L^{}} \gamma^{}_{ti}} .
\end{eqnarray}

Once the model parameters have been fitted by iteration of the above update procedure to convergence (which may only converge to a local maximum of the likelihood), the most likely hidden state sequence can be determined given the observations $\bfm{O}$ and the MLE model $\hat{\bfm{\Theta}}$ using the Viterbi algorithm~\cite{viterbi:1967:IEEE-trans-info-theory:viterbi-algorithm}.
Like the forward-backward algorithm employed in the Baum-Welch procedure, the Viterbi algorithm also has a forward recursion component,
\begin{eqnarray}
\epsilon_{jt} &=& \begin{cases}
\rho_j \varphi(o_t \mid \bfm{e}_j) & t = 0 \\
\varphi(o_t \mid \bfm{e}_j) \max_{i} \epsilon_{i(t-1)} T_{ij}  & t = 1,\ldots,L
\end{cases} \label{equation:viterbi-forward} \\
\Phi_{jt} &=& \begin{cases}
1 & t = 0 \\
\arg\max_{i} \epsilon_{i(t-1)} T_{ij}  & t = 1,\ldots,L
\end{cases} \nonumber
\end{eqnarray}
as well as a reverse reconstruction component to compute the most likely state sequence $\hat{\bfm{S}}$,
\begin{eqnarray}
\hat{s}_t &=& \begin{cases}
\arg\max_i \epsilon_{it} & t = L\\
\Phi_{\hat{s}_{t+1}(t+1)} & t = (L-1),\ldots,0
\end{cases} 
\end{eqnarray}

\subsection{Sampling from the posterior of the BHMM}
\label{section:algorithms:bhmm-posterior-sampling}

Sampling from the posterior of the BHMM (Eq.~\ref{equation:full-posterior}) proceeds by rounds of Gibbs sampling, where each round consists of an update of the augmented model parameters $\{\bfm{T}, \bfm{E}, \bfm{S}\}$ by sampling
\begin{eqnarray}
\begin{array}{lcl}
\bfm{S}' \mid \bfm{T}, \bfm{E}, \bfm{O} &\sim& P(\bfm{S}' \mid \bfm{T}, \bfm{E}, \bfm{O}) \nonumber \\
\bfm{T}' \mid \bfm{S}' &\sim& P(\bfm{T}' \mid \bfm{S}') \nonumber \\
\bfm{E}' \mid \bfm{S}', \bfm{O} &\sim& P(\bfm{E}' \mid \bfm{S}', \bfm{O}) 
\end{array}
\end{eqnarray}
where the conditional probabilities are given by Eq.~\ref{equation:conditional-gibbs}.  

\subsubsection{Updating the hidden state sequences}

We use a modified form of the Viterbi process to generate an independent sample of the hidden state history $\bfm{S}$ given the transition probabilities $\bfm{T}$, state observable distribution parameters $\bfm{E}$, and observed data $\bfm{O}$.
Like the Viterbi scheme, a forward recursion is applied to each observation trace $\bfm{o}^{}$, but instead of computing the most \emph{likely} state history on the reverse pass, a new hidden state history $\bfm{S}$ is drawn from the distribution $P(\bfm{S} \mid \bfm{O}, \bfm{T}, \bfm{E})$.
The forward recursion uses the same forward algorithm as used in Baum-Welch~\cite{baum:1970:ann-math-statist:baum-welch}, 
\begin{eqnarray}
\alpha_{tj} &=& \begin{cases}
\rho_j \, \varphi(o_0 \mid \bfm{e}_j) & t = 0 \\
\varphi(o_t \mid \bfm{e}_j) \sum_{i=1}^M \alpha_{(t-1)i} T_{ij} & t = 1,\ldots,L
\end{cases}
\end{eqnarray}
In the reverse recursion, we now \emph{sample} a state sequence by sampling each hidden state from the conditional distribution $s_t \sim P(s_t \mid s_{t+1}, \ldots, s_L)$ starting from $t = L$ and proceeding down to $t = 0$, where the conditional distribution is given by,
\begin{eqnarray}
\lefteqn{P(s_t = i \mid s_{t+1}, \ldots, s_L)} \\
&\propto& \begin{cases}
\alpha_{t i} / \sum_{j=1}^M \alpha_{t j} & t = L \\
\alpha_{t i} T_{i s_{t+1}} / \sum_{j=1}^M \alpha_{t j} T_{j s_{t+1}} & t = (L-1),\ldots,0
\end{cases} \nonumber
\end{eqnarray} 
It is straightforward to show the result of these sampling steps reconstitutes the probability distribution $P(\bfm{S} | \bfm{T}, \bfm{E}, \bfm{O})$ (see \emph{Supplementary Material: Proof of state history sampling scheme}).

\subsubsection{Updating the transition probabilities}

If no detailed balance constraint is used and the prior $P(\bfm{T})$ is Dirichlet in each row of the transition matrix $\bfm{T}$, it is possible to generate an independent sample of the transition matrix from the conditional distribution $P(\bfm{T}' \mid \bfm{S}')$ by sampling each row of the transition matrix from the conjugate Dirichlet posterior using the transition counts from the sampled state sequence $\bfm{S}'$~\cite{noe:jcp:2008:transition-matrix-sampling}.
However, because physical systems in the absence of energy input through an external driving force should satisfy detailed balance, we make use of this constraint in updating our transition probabilities, since this has been demonstrated to substantially reduce parameter uncertainty in the data-limited regime~\cite{noe:jcp:2008:transition-matrix-sampling}.

The transition matrix is updated using the reversible transition matrix sampling scheme of No\'{e}~\cite{noe:jcp:2008:transition-matrix-sampling,chodera-noe:jcp:2011:observable-sampling}.
Here, an adjusted count matrix $\bfm{C} \equiv (c_{ij})$ is computed using the updated hidden state sequence $\bfm{S}'$,
\begin{eqnarray}
c_{ij} &=& b_{ij} +\sum_{t=1}^{L} \delta_{is_{t-1}} \delta_{js_{t}} ,
\end{eqnarray}
where the Kronecker $\delta_{ij} = 1$ if $i = j$ and zero otherwise, and $\bfm{B} \equiv (b_{ij})$ is a matrix of prior pseudocounts, which we take to be zero following the work of No\'{e} et al.~\cite{noe:pnas:2009:ww-domain}.
Using the adjusted count matrix $\bfm{C}$, a Metropolis-Hastings Monte Carlo procedure~\cite{metropolis-hastings} is used to update the matrix and produce a new sample from $P(\bfm{T}' \mid \bfm{S}')$.
Two move types are attempted, selected with equal probability, and 1000 moves are attempted to generate a new sample $\bfm{T}'$ that is approximately uncorrelated from the previous $\bfm{T}$.
Prior to starting the Monte Carlo procedure, the vector of equilibrium probabilities for all states $\bfv{\pi}$ is computed according to 
\begin{eqnarray}
\bfm{T}^\T \bfv{\pi} = \bfv{\pi} .
\end{eqnarray}

The first move type is a \emph{reversible element shift}.
A pair of states $(i,j)$, $i \ne j$, are selected with uniform probability, and a random number $\Delta$ is selected uniformly over the interval,
\begin{equation*}
\Delta \in[\max(-T_{ii},-\frac{\pi_{j}}{\pi_{i}}T_{jj}),T_{ij}] .
\end{equation*}
The changed elements in the proposed transition matrix $\bfm{T}'$ are then given by:
\begin{align*}
T'_{ij} & =T_{ij}-\Delta \:\: ; \:\: T'_{ji}  =T_{ji}-\frac{\pi_{i}}{\pi_{j}}\Delta \\
T'_{ii} & =T_{ii}+\Delta \:\: ; \:\: T'_{jj}  =T_{jj}+\frac{\pi_{i}}{\pi_{j}}\Delta  .
\end{align*}
This move is accepted with probability
\begin{eqnarray}
\lefteqn{P_\mathrm{accept}(\bfm{T}' | \bfm{T} ) = \mathrm{min}\left\{1, \sqrt{\frac{(T'_{ij})^{2}+(T'_{ji})^{2}}{(T_{ij})^{2}+(T_{ji})^{2}}} \right. }\\
&\times& \left. \left(\frac{T'_{ii}}{T_{ii}}\right)^{c_{ii}}\left(\frac{T'_{ij}}{T_{ij}}\right)^{c_{ij}} \left(\frac{T'_{jj}}{T_{jj}}\right)^{c_{jj}}\left(\frac{T'_{ji}}{T_{ji}}\right)^{c_{ji}} \right\} . \nonumber
\end{eqnarray}
This move will leave the vector of stationary probabilities $\bfv{\pi}$ unchanged.

The second move type is a \emph{row shift}.
A row $i$ of $\bfm{T}$ is selected with uniform probability, and a random number $\eta$ chosen uniformly over the interval 
\begin{equation*}
\eta\in\left[0,\frac{1}{1-T_{ii}}\right]
\end{equation*}
and used to update row $i$ of $\bfm{T}$ according to
\begin{eqnarray}
T'_{ij} &=& \begin{cases}
\eta T_{ij} & j = 1, \ldots, M, \:\: j \ne i\\
\eta (T_{ii} - 1) + 1 & j = i
\end{cases}
\end{eqnarray}
This move is accepted with probability
\begin{eqnarray}
P_\mathrm{accept}(\bfm{T}' | \bfm{T}) &=& \mathrm{min} \left\{1, \eta^{(M-2)} \eta^{(c_{i*}-c_{ii})}\left(\frac{1-\eta(1-T_{ii})}{T_{ii}}\right)^{c_{ii}} \right\} . \nonumber\\
\end{eqnarray}
The row shift operation will change the stationary distribution of $\bfv{\pi}'$, but it may be efficiently updated with
\begin{align*}
\pi'_{i} & =\frac{\pi_{i}}{\pi_{i}+\eta(1-\pi_{i})} \:\: ; \:\: \pi'_{j} =\frac{\eta \, \pi_{j}}{\pi_{i}+\eta(1-\pi_{i})}.
\end{align*}
Since this update scheme is incremental, it will accumulate numerical errors over time that cause the updated $\bfv{\pi}$ to drift away from the stationary distribution of the current transition matrix. 
To avoid this, $\bfv{\pi}$ is recomputed from the current sample of the transition matrix in regular intervals (here, every 100 sampling steps).

\subsubsection{Updating the observable distribution parameters}

Following the update of the transition matrix $\bfm{T}$, the observable distribution parameters $\bfm{E}$ are updated by sampling $\bfm{E}$ from the conditional probability $P(\bfm{E}' \mid \bfm{S}', \bfm{O})$.
The conditional probability for the observable distribution parameters for state $m$, denoted $\bfm{e}_m$, is given in terms of the output model $\varphi(o \mid \bfm{e})$ by Bayes' theorem,
\begin{eqnarray}
P(\bfm{E} \mid \bfm{O}, \bfm{S}) &=& \left[\prod_{t=0}^{L^{}} \varphi(o_{t} \mid \bfm{e}_{s^{}_{t}}) \right] P(\bfm{E}) \label{equation:conditional-emission-probability} .
\end{eqnarray}

An important choice must be made with regards to the prior, $P(\bfm{E})$.
If the prior is chosen to be composed of independent priors for each state, as in
\begin{eqnarray}
P(\bfm{E}) &=& \prod_{m=1}^M P(\bfm{e}_m) \label{equation:separable-emission-prior} ,
\end{eqnarray}
then the full BHMM posterior (Eq.~\ref{equation:full-posterior}) will be invariant under any permutation of the states.
This behavior might be undesirable, as the states may switch labels during the posterior sampling procedure; this will require any analysis of the models sampled from the posterior to account for the possible permutation symmetry in the states.
On the other hand, breaking this symmetry (e.g., by enforcing an ordering on the state mean observables) can artificially restrict the confidence intervals of the states, which might additionally complicate data analysis.

Here, we make the choice that the prior be separable (Eq.~\ref{equation:separable-emission-prior}), which has the benefit of allowing the conditional probability for $\bfm{E}$ (Eq.~\ref{equation:conditional-emission-probability}) to be decomposed into a separate posterior for each state.
For each state $m$, collect all the observations $o^{}_{t}$ whose updated hidden state labels ${s^{}_{t}}' = m$ into a single dataset $\bfm{o} \equiv \{o_n\}_{n=1}^{N_m}$, where $N_m$ is the total number of times state $m$ is visited, for the purposes of this update procedure.
Then, the observable parameters $\bfm{e}$ for this state are given by
\begin{eqnarray}
P(\bfm{e} \mid \bfm{o}) &=& P(\bfm{o} \mid \bfm{e}) P(\bfm{e}) = \left[ \prod_{n=1}^{N_m} \varphi(o_n \mid \bfm{e}) \right] P(\bfm{e}) .
\end{eqnarray}

In the application presented here, we use a Gaussian output model (Eq.~\ref{equation:gaussian-observable}) 
for the state observable distributions $P(o \mid \bfm{e})$, where $\bfm{e} \equiv \{\mu, \sigma^2\}$, with $\mu$ the state mean observable and $\sigma^2$ the variance (which will include both the distribution of the observable characterizing the state and any broadening from measurement noise).
Other models (including multidimensional or multimodal observation models) are possible, and require replacing only the observation model $\varphi(o \mid \bfm{e})$ and corresponding prior $P(\bfm{e})$.

We use the (improper) Jeffreys prior~\cite{jeffreys:proc-royal-soc-a:1946:jeffreys-prior} which has the information-theoretic interpretation as the prior that maximizes the information content of the data~\cite{goyal:bayesian-inference-book:2005:jeffreys-prior}, (suppressing the state index subscript $m$),
\begin{eqnarray}
P(\bfm{e}) &\propto& \sigma^{-1} \label{equation:emission-parameter-prior} ,
\end{eqnarray}
which produces the posterior
\begin{eqnarray}
P(\bfm{e} \mid \bfm{o}) &\propto& \sigma^{-(N+1)} \exp\left[-\frac{1}{2 \sigma^2} \sum_{n=1}^N (o_n - \mu)^2 \right] ,
\end{eqnarray}
where we remind the reader that here and in the remainder of this section, the symbols $\bfm{e}$, $\bfm{o}$, $\sigma$, $\mu$, and $N$ refer to $\bfm{e}_m$, $\bfm{o}_m$, $\sigma_m$, $\mu_m$, and $N_m$, respectively.

Updating $\{\mu, \sigma^2\}$ also proceeds by a Gibbs sampling scheme, alternately updating $\mu$ and $\sigma$, as earlier described in Ref.~\citep{chodera-noe:jcp:2011:observable-sampling},
\begin{eqnarray}
\mu &\sim& P(\mu \mid \sigma^2, \bfm{o}) \nonumber \\
\sigma^2 &\sim& P(\sigma^2 \mid \mu, \bfm{o})
\end{eqnarray}

The conditional distribution of the mean $\mu$ is then given by
\begin{eqnarray}
P(\mu \mid \sigma^2, \bfm{o}) &\propto& \exp\left[-\frac{1}{2 (\sigma^2/N)} (\mu - \hat{\mu})^2 \right] 
\end{eqnarray}
where $\hat{\mu}$ is the sample mean for $\bfm{o}$, the samples in state $m$,
\begin{eqnarray}
\hat{\mu} &\equiv& \frac{1}{N} \sum_{n=1}^N o_n
\end{eqnarray}
This allows us to update $\mu$ according to 
\begin{eqnarray}
\mu' &\sim& \mathcal{N}(\hat{\mu}, \sigma^2 / N)
\end{eqnarray}

The conditional distribution of the variance $\sigma^2$ is given by
\begin{eqnarray}
P(\sigma^2 \mid \mu, \bfm{o}) &\propto& \sigma^{-(N+1)} \exp\left[-\frac{N \hat{\sigma}^2}{2 \sigma^2} \right] 
\end{eqnarray}
where the quantity $\hat{\sigma}^2$, which is \emph{not} in general identical to the sample variance, is given by
\begin{eqnarray}
\hat{\sigma}^2 &\equiv& \frac{1}{N} \sum_{n=1}^N (o_n - \mu)^2 .
\end{eqnarray}
A convenient way to update $\sigma^2 \mid \mu, \bfm{o}$ is to sample a random variate $y$ from the chi-square distribution with $N-1$ degrees of freedom,
\begin{eqnarray}
y &\sim& \chi^2(N-1)
\end{eqnarray}
and then update $\sigma^2$ as
\begin{eqnarray}
{\sigma'}^2 &=& N \hat{\sigma}^2 / y .
\end{eqnarray}
Note that $\mu$ and $\sigma^2$ can be updated in either order, but the \emph{updated} values of $\mu$ or $\sigma^2$ must be used in sampling the not-yet-updated $\sigma^2$ or $\mu$, and vice-versa.

Other output probabilities, such as \emph{mixtures} of normal distributions or other distributions, can be substituted by simply changing $P(\bfm{E} \mid \bfm{O}, \bfm{S})$ and the scheme by which $\bfm{E}$ is updated.


\begin{figure*}[tbp]
\noindent
\resizebox{\textwidth}{!}{\includegraphics{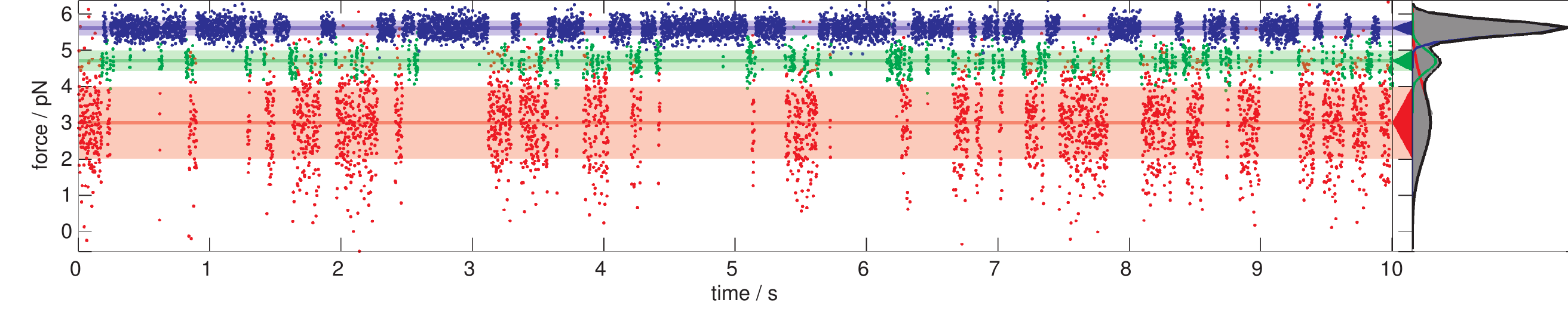}}
\caption{{\bf Synthetic force trajectory and inferred state assignments in MLHMM.}
Observed samples are colored by their hidden state assignments.
Dark horizontal lines terminating in triangles to the right denote state means, while lightly colored bands indicate one standard deviation on either side of the state mean.
The gray histogram on the right side shows the total observed probability of samples, while the colored peaks show the weighted Gaussian output contribution from each state, and the black outline the weighted sum of the Gaussian output contributions from the HMM states.
\label{figure:model-stateassignments}}
\end{figure*}

\begin{table*}
\caption{{\bf Estimated mean model parameters and confidence intervals for synthetic timeseries data}}
\label{table:synthetic-confidence-intervals}
\begin{tabular*}{\textwidth}{@{\extracolsep{\fill}}lccccc}
\hline
&  &  & \multicolumn{3}{c}{\bf Estimated Model Parameters}  \\ \cline{4-6}
\multicolumn{2}{l}{\bf Property} & \bf True Value & \bf 1 000 observations & \bf 10 000 observations & \bf 100 000 observations\\ \hline
stationary probability & $\pi_{1}$ & $0.308$ & $0.228_{\:0.074}^{\:0.480}$ & $0.318_{\:0.244}^{\:0.407}$ & $0.324_{\:0.292}^{\:0.355}$ \\
& $\pi_{2}$ & $0.113$ & $0.093_{\:0.042}^{\:0.172}$ & $0.124_{\:0.098}^{\:0.155}$ & $0.112_{\:0.104}^{\:0.121}$ \\
& $\pi_{3}$ & $0.579$ & $0.679_{\:0.415}^{\:0.870}$ & $0.558_{\:0.455}^{\:0.648}$ & $0.564_{\:0.531}^{\:0.599}$ \\
\hline
transition probability & $T_{11}$ & $0.980$ & $0.970_{\:0.945}^{\:0.987}$ & $0.972_{\:0.966}^{\:0.978}$ & $0.979_{\:0.978}^{\:0.981}$ \\
& $T_{12}$ & $0.019$ & $0.023_{\:0.009}^{\:0.045}$ & $0.026_{\:0.021}^{\:0.032}$ & $0.020_{\:0.018}^{\:0.021}$ \\
& $T_{13}$ & $0.001$ & $0.007_{\:0.001}^{\:0.018}$ & $0.002_{\:0.001}^{\:0.003}$ & $0.001_{\:0.001}^{\:0.001}$ \\
& $T_{21}$ & $0.053$ & $0.054_{\:0.018}^{\:0.106}$ & $0.067_{\:0.053}^{\:0.082}$ & $0.057_{\:0.052}^{\:0.061}$ \\
& $T_{22}$ & $0.900$ & $0.868_{\:0.790}^{\:0.931}$ & $0.890_{\:0.870}^{\:0.907}$ & $0.897_{\:0.892}^{\:0.903}$ \\
& $T_{23}$ & $0.050$ & $0.078_{\:0.035}^{\:0.136}$ & $0.043_{\:0.033}^{\:0.056}$ & $0.046_{\:0.042}^{\:0.050}$ \\
& $T_{31}$ & $0.001$ & $0.002_{\:0.000}^{\:0.006}$ & $0.001_{\:0.000}^{\:0.002}$ & $0.001_{\:0.000}^{\:0.001}$ \\
& $T_{32}$ & $0.009$ & $0.010_{\:0.004}^{\:0.019}$ & $0.010_{\:0.007}^{\:0.012}$ & $0.009_{\:0.008}^{\:0.010}$ \\
& $T_{33}$ & $0.990$ & $0.988_{\:0.978}^{\:0.995}$ & $0.990_{\:0.987}^{\:0.992}$ & $0.990_{\:0.989}^{\:0.991}$ \\
\hline
state mean force (pN) & $\mu_{1}$ & $3.000$ & $2.947_{\:2.812}^{\:3.082}$ & $2.998_{\:2.963}^{\:3.033}$ & $3.001_{\:2.990}^{\:3.013}$ \\
& $\mu_{2}$ & $4.700$ & $4.666_{\:4.612}^{\:4.721}$ & $4.699_{\:4.683}^{\:4.716}$ & $4.702_{\:4.696}^{\:4.707}$ \\
& $\mu_{3}$ & $5.600$ & $5.597_{\:5.583}^{\:5.614}$ & $5.602_{\:5.596}^{\:5.607}$ & $5.602_{\:5.600}^{\:5.603}$ \\
\hline
state std dev force (pN) & $\sigma_{1}$ & $1.000$ & $1.037_{\:0.951}^{\:1.134}$ & $0.992_{\:0.967}^{\:1.018}$ & $0.999_{\:0.991}^{\:1.007}$ \\
& $\sigma_{2}$ & $0.300$ & $0.254_{\:0.217}^{\:0.300}$ & $0.287_{\:0.275}^{\:0.300}$ & $0.301_{\:0.296}^{\:0.305}$ \\
& $\sigma_{3}$ & $0.200$ & $0.200_{\:0.190}^{\:0.211}$ & $0.203_{\:0.199}^{\:0.207}$ & $0.201_{\:0.200}^{\:0.203}$ \\
\hline
\end{tabular*}
\end{table*}


\section{Validation using synthetic data}
\label{section:model-system}

To verify that our BHMM posterior sampling scheme reflects the \emph{true} uncertainty in the model parameters, we tested the scheme on synthetic data generated from a model with known parameters $\bfm{\Theta}^*$. 
Given observed data $\bfm{O}$ generated from $P(\bfm{O} \mid \bfm{\Theta}^*)$, sampling from the posterior $P(\bfm{\Theta} \mid \bfm{O})$ using the scheme described in \emph{Sampling from the posterior of the BHMM} will provide us with confidence intervals $[\theta_\mathrm{low}, \theta_\mathrm{high}]$ for a specified confidence interval size $\alpha \in [0,1]$.
If these computed confidence intervals are accurate, we should find that the true model parameter $\theta^*$ lies in the computed confidence interval $[\theta^{(\alpha)}_\mathrm{low}, \theta^{(\alpha)}_\mathrm{high}]$ with probability $\alpha$.
This can be tested by generating synthetic observed data $\bfm{O}$ from $P(\bfm{O} \mid \bfm{\Theta}^*)$ and verifying that we find $\theta^* \in [\theta^{(\alpha)}_\mathrm{low}, \theta^{(\alpha)}_\mathrm{high}]$ in a fraction $\alpha$ of these synthetic experiments.

As an example synthetic model, consider the three-state system intended to mimic a protein with (1) a highly-compliance, low-force unfolded state, (2) a moderately compliant low-population intermediate at intermediate force, and (3) a low-compliance, high-force folded state.
Here, the term ``compliance'' refers to the width of the force or extension distribution characterizing the state.
Parameters of the model are given in Table~\ref{table:synthetic-confidence-intervals}, and the observation interval was taken to be $\tau = 1$ ms.
An example realization of a model trajectory, along with the MLHMM state assignment, is shown in Figure~\ref{figure:model-stateassignments}.
We generated a trajectory of 100 000 observations, and characterized the BHMM mean parameter estimate and 95\% confidence intervals for a subset of this trajectory of varying lengths.
The results, shown in Table~\ref{table:synthetic-confidence-intervals}, show that the confidence intervals contract as trajectory length increases, as expected, and the BHMM-computed 95\% confidence intervals contain the true model parameters with the expected statistics.
In contrast, a model created from simply segmenting the observed forces into disjoint region and assigning state membership based on the force value alone estimates model parameters with significant bias even for 1 000 000 observations (see \emph{Supporting Information}).

As a more rigorous test, we sampled 50 random models from the prior $P(\bfm{\Theta})$ with two to six states, generated a 10 000 observation synthetic trajectory for each, and accumulated statistics on the observed fraction of time the true model parameters were within the BHMM confidence intervals for various values of the confidence interval width $\alpha$.
The results of this test are depicted in \emph{Supplementary Figure 1}. 
We expect that the plot traces the diagonal if the observed and expected confidence intervals are identical; an overestimate of the confidence interval will be above the diagonal, and an underestimate will fall below it.
Because only a finite number of independent replicates of the experiment are conducted, there is some associated uncertainty with the observed confidence intervals.
The results show that the observed confidence intervals line up with the expected confidence intervals to within statistical error, suggesting the BHMM confidence intervals neither underestimate nor overestimate the actual uncertainty in model parameters.


\begin{figure*}[tbp]
\noindent
\resizebox{\textwidth}{!}{\includegraphics{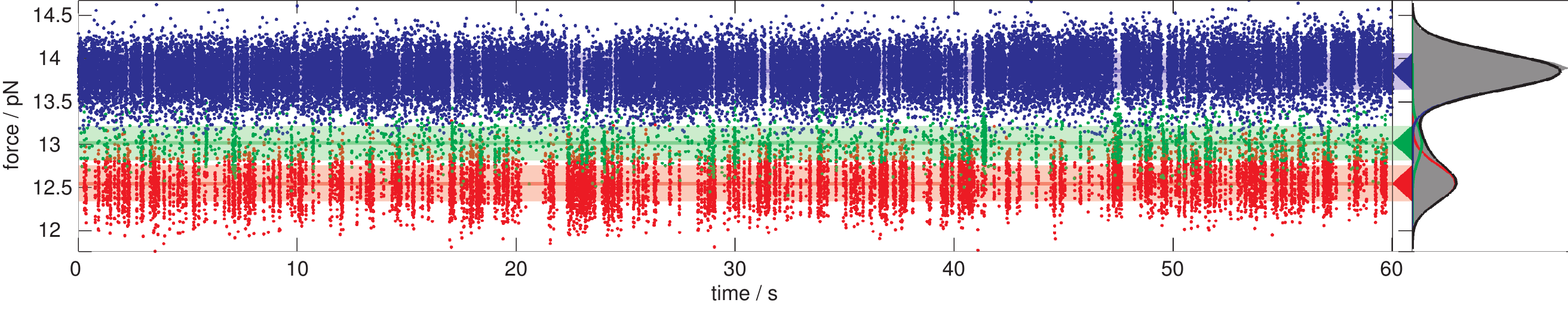}}
\caption{{\bf Experimental force trajectory of the p5ab hairpin and MLHMM state assignments.}
Observed samples are colored by their hidden state assignments.
Dark horizontal lines terminating in triangles to the right denote state means, while lightly colored bands indicate one standard deviation on either side of the state mean.
The gray histogram on the right side shows the total observed probability of samples, while the colored peaks show the weighted Gaussian output contribution from each state, and the black outline the weighted sum of the Gaussian output contributions from the HMM states.
\label{figure:p5ab-observed-trajectory}}
\end{figure*}

\begin{table}
\caption{{\bf BHMM model estimates for p5ab hairpin data.}
}
\label{table:p5ab-bhmm-confidence-intervals}
\begin{tabular*}{\columnwidth}{@{\extracolsep{\fill}}lcc}
\hline
\multicolumn{2}{l}{\bf Property} & \bf Value\\ \hline
Equilibrium probability & $\pi_{1}$ & $0.215_{\:0.193}^{\:0.236}$ \\
& $\pi_{2}$ & $0.046_{\:0.041}^{\:0.050}$ \\
& $\pi_{3}$ & $0.740_{\:0.717}^{\:0.762}$ \\
\hline
Transition probability ($\Delta t = 1$ ms) & $T_{11}$ & $0.954_{\:0.950}^{\:0.959}$ \\
& $T_{12}$ & $0.033_{\:0.029}^{\:0.037}$ \\
& $T_{13}$ & $0.013_{\:0.011}^{\:0.015}$ \\
& $T_{21}$ & $0.154_{\:0.139}^{\:0.169}$ \\
& $T_{22}$ & $0.650_{\:0.627}^{\:0.673}$ \\
& $T_{23}$ & $0.196_{\:0.180}^{\:0.216}$ \\
& $T_{31}$ & $0.004_{\:0.003}^{\:0.004}$ \\
& $T_{32}$ & $0.012_{\:0.011}^{\:0.013}$ \\
& $T_{33}$ & $0.984_{\:0.983}^{\:0.985}$ \\
\hline
State force mean (pN) & $\mu_{1}$ & $12.549_{\:12.544}^{\:12.552}$ \\
& $\mu_{2}$ & $13.016_{\:13.006}^{\:13.027}$ \\
& $\mu_{3}$ & $13.849_{\:13.848}^{\:13.852}$ \\
\hline
State force std dev (pN) & $\sigma_{1}$ & $0.210_{\:0.207}^{\:0.213}$ \\
& $\sigma_{2}$ & $0.201_{\:0.193}^{\:0.208}$ \\
& $\sigma_{3}$ & $0.213_{\:0.211}^{\:0.214}$ \\
\hline \hline
Transition rate (s$^{-1}$) & $k_{12}$ & $41.4_{\:36.3}^{\:46.6}$ \\
& $k_{13}$ & $9.1_{\:7.2}^{\:11.3}$ \\
& $k_{21}$ & $194.7_{\:173.1}^{\:216.7}$ \\
& $k_{23}$ & $243.7_{\:219.0}^{\:271.5}$ \\
& $k_{31}$ & $2.6_{\:2.1}^{\:3.2}$ \\
& $k_{32}$ & $15.0_{\:13.4}^{\:16.6}$ \\
\hline
State mean lifetime (ms) & $\tau_{1}$ & $21.9_{\:20.0}^{\:24.1}$ \\
& $\tau_{2}$ & $2.9_{\:2.7}^{\:3.1}$ \\
& $\tau_{3}$ & $63.1_{\:58.4}^{\:68.5}$ \\
\hline
\end{tabular*}
\end{table}

\section{RNA hairpin kinetics in a passive optical trap}
\label{section:application}

We illustrate the BHMM approach applied to real force spectroscopy data by characterizing the average forces and transition rates among kinetically distinct states of the p5ab RNA hairpin in an optical trap under passive (equilibrium) conditions.

The p5ab RNA hairpin from \emph{Tetrahymena thermophilia} was provided by Jin-Der Wen, and prepared as previously described~\cite{wen-manosas:biophys-j:2007:rna-optical-tweezers}.  
Within the population of RNA hairpin molecules in the examined sample, there were two chemically distinct species present in the sample (i.e.~as a result of post-transcriptional or other covalent modification during sample storage), exhibiting either apparent two-state (as reported previously~\cite{wen-manosas:biophys-j:2007:rna-optical-tweezers}) or three-state behavior (studied here).  
For the purposes of testing this method, we examined a fiber that appeared to consistently exhibit three-state behavior upon visual inspection of the force timeseries data. 

The instrument used in this experiment was a dual-beam counter-propagating optical trap with a spring constant of 0.1 pN/nm.  
A piezoactuator controlled the position of the trap and allowed position resolution to within 0.5 nm~\cite{bustamante-smith:2006:minitweezers-patent}.  
Drift in the instrument was less than 1 nm/minute resulting in a constant average force within 0.1 pN over the course of a typical 60 s experiment.  
For these constant trap position experiments, higher frequency data was recorded at 50 kHz recording the voltage corresponding to the force on the tether directly from the position-sensitive detectors.
To ensure sequential samples obeyed Markovian statistics, these data were subsampled down to 1 kHz for analysis by the BHMM framework after examination of autocorrelation functions for trap positions where the hairpin appeared to remain in a single conformational state (see \emph{Supplementary Material: Choice of observation interval}).

A single observed force trajectory at a fixed trap position adequate to cause hopping among multiple states is shown in Figure~\ref{figure:p5ab-observed-trajectory}.
The most likely state trajectory from the MLHMM fit with three states is shown by coloring the observations most likely to be associated with each state, with bands of color indicating the mean and standard deviation about the mean force characterizing each state.

Table~\ref{table:p5ab-bhmm-confidence-intervals} lists the BHMM posterior means and confidence intervals characterizing the three-state model extracted from this single 60 s observed force trace.
Several things are notable about the estimated model parameters.
Surprisingly, while there is a clearly-resolved intermediate-force state (\emph{state 2}) through which most of the flux from the high- and low-force states passes (as seen from large $K_{12}$ and $K_{23}$), there are nontrivial rate constants connecting the high and low force states directly ($K_{13}$), indicating that while a sequential mechanism involving passing through the intermediate state is preferred, it may not be an obligatory step in hairpin formation under these conditions.
While the state mean forces are clearly distinct, the state standard deviations---which reflect the width of the observed force distribution characterizing each state, rather than the uncertainty in state means---possess overlapping confidence intervals.
These standard deviations reflect not only contributions from both the distribution of extensions sampled by the hairpin in each conformational state, but also from fluctuations in the handles and beads, and other sources of mechanical and electrical noise in the measurement.
As we would expect the unfolded hairpin to be more compliant (i.e.~possess a wider distribution of forces) than the folded hairpin, the inability to distinguish the standard deviations among states is suggestive that, for this experimental configuration and observation time, the predominant contribution to the observed force distributions for each state may be in the form of handle or bead fluctuations or other sources of measurement noise.

Finally, the lifetime of the intermediate-force state is significantly shorter than for the low- and high-force states by nearly an order of magnitude, and only a few times longer than the observation interval of 1 ms---despite this, the lifetime appears to be well-determined, as indicated by the narrow confidence intervals.

\section{Discussion}
\label{section:discussion}

We have described an approach to determining the first-order kinetic parameters and observable (force or extension) distributions characterizing conformational states in single-molecule force spectroscopy.
By use of a Bayesian extension of hidden Markov models, we are able to characterize the experimental uncertainty in these parameters due to instrument noise and finite-size datasets.
The use of a detailed balance constraint additionally helps reduce the experimental uncertainty over standard hidden Markov models, as both transitions into and out of conformational states provide valuable information about state kinetics and populations in data-poor conditions~\cite{noe:jcp:2008:transition-matrix-sampling,metzner-noe-schuette:pre:2009:transition-matrix-sampling}.
Additionally, the Gibbs sampling framework used to sample from the Bayesian posterior can be easily extended to incorporate additional nuisance parameters, such as stochastic models of instrument drift or laser power fluctuations.

We have opted to make use of a reversible transition matrix to describe the statistical kinetic behavior between the observation intervals $\Delta t$, but it is possible to use a reversible rate matrix instead by substituting a rate matrix sampling scheme~\cite{hummer:njp:2005:rate-matrix-sampling} in the appropriate stage of the Gibbs sampling updates.

While the experimenter must currently choose the number of conformational states by hand, a number of extensions of Bayesian hidden Markov models can be used to automatically determine the number of states best supported by the data, including reversible-jump schemes~\cite{robert:j-r-statist-soc-b:2000:reversible-jump-mcmc,degunst-schouten:bernoulli:2003:reversible-jump-hmm} and variational Bayes methods~\cite{beal:variational-bayes,bronson:biophys-j:2009:variational-bayesian-hmm}.

We note that the experimenter in principle has access to the full posterior distribution of models given the observed data, so that instead of looking at the confidence of single parameters, confidence intervals in more complex functions of parameters---such as the rates or lifetimes in Table~\ref{table:p5ab-bhmm-confidence-intervals}---can be computed, or joint posterior distributions of multiple parameters examined.
It is also possible to generate \emph{synthetic} data from the current model, or family of models, to examine how the collection of additional data will further reduce uncertainties or allow discrimination among particular hypotheses.
The field of \emph{Bayesian experimental design}~\cite{chaloner:stat-sci:1995:bayesian-experimental-design-review} holds numerous possibilities for selecting how future experiments can maximize information gain, and whether the information gain from the collection of additional data will be of sufficient utility to justify the expense.


\section{Acknowledgments}
The authors thank Sergio Bacallado (Stanford University) for helpful feedback on this manuscript, and Steve Presse (UCSF) for engaging discussions on this topic.
JDC acknowledges support from a QB3-Berkeley Distinguished Postdoctoral Fellowship.
FN acknowledges DFG Grant 825/2.
This work was supported in part by a grant from the NSF (SM).


\bibliography{bhmm}

\end{document}